\documentclass[12pt] {article}
\usepackage{psfrag}
\usepackage{graphicx}
\usepackage{latexsym,amsfonts}
\usepackage{amssymb}
\begin{document}
\setcounter{page}{1}
\def\theequation{\arabic{section}.\arabic{equation}}
\def\theequation{\thesection.\arabic{equation}}
\setcounter{section}{0}

\title{Energy level displacement of the excited $n\ell$ state of
pionic hydrogen}

\author{A. N. Ivanov\,\thanks{E--mail: ivanov@kph.tuwien.ac.at, Tel.:
+43--1--58801--14261, Fax: +43--1--58801--14299}~\thanks{Permanent
Address: State Polytechnical University, Department of Nuclear
Physics, 195251 St. Petersburg, Russian Federation}\,,
M. Faber\,\thanks{E--mail: faber@kph.tuwien.ac.at, Tel.:
+43--1--58801--14261, Fax: +43--1--58801--14299},
A. Hirtl\,\thanks{E--mail: albert.hirtl@oeaw.ac.at}\,,
J. Marton\,\thanks{E--mail: johann.marton@oeaw.ac.at}\,,
N. I. Troitskaya\,\thanks{State Polytechnical University, Department
of Nuclear Physics, 195251 St. Petersburg, Russian Federation}}

\date{\today}

\maketitle

\vspace{-0.5in}
\begin{center}
{\it Atominstitut der \"Osterreichischen Universit\"aten,
Arbeitsbereich Kernphysik und Nukleare Astrophysik, Technische
Universit\"at Wien, \\ Wiedner Hauptstr. 8-10, A-1040 Wien,
\"Osterreich }
\end{center}

\begin{center}
\begin{abstract}
The energy level displacements of the excited $n\ell$ states of pionic
hydrogen and the contribution of the $ns \to 1s$ transitions and the
$(\pi^-p)_{\rm Coul} \to 1s$ transitions of the $\pi^-p$ pair, coupled
by the attractive Coulomb field in the S--wave state with a continuous
energy spectrum, to the shift of the energy level of the ground state
of pionic hydrogen, caused by strong low--energy interactions, are
calculated within a quantum field theoretic, relativistic covariant
and model--independent approach developed in nucl--th/0306047.
\end{abstract}
\end{center}

\newpage

\section{Introduction}

Pionic hydrogen $A_{\pi p}$, a hydrogen--like mesoatom with the
electron replaced by the $\pi^-$ meson, is a nice laboratory for the
experimental investigation of strong low--energy interactions and the
mechanisms of spontaneous breaking of chiral symmetry
\cite{PSI1,PSI2}.

As has been found by Deser, Goldberger, Baumann and Thirring
\cite{SD54} (see also \cite{TT61}--\cite{TE03} and the textbook by
Ericson and Weise \cite{TE88}) due to strong low--energy interactions
the energy level of the ground state of pionic hydrogen acquires the
following shift and width
\begin{eqnarray}\label{label1.1}
- \epsilon_{1s} + i\,\frac{\Gamma_{1s}}{2}&=&
\frac{2\pi}{\mu}\,\Big[\frac{1}{3}\,(2a^{1/2}_0 + a^{3/2}_0) + i\,
\frac{2}{9}\,Q_{\pi^0n}\,(a^{1/2}_0 -
a^{3/2}_0)^2\Big]\,|\Psi_{1s}(0)|^2=\nonumber\\ &=&
\frac{2\pi}{\mu}\,f^{\pi^-p}_0(0)\,|\Psi_{1s}(0)|^2.
\end{eqnarray}
This is the so--called DGBT formula, $f^{\pi^-p}_0(0)$ is the S--wave
amplitude of low--energy $\pi^-p$ scattering calculated at threshold,
$a^{^1/2}_0$ and $a^{3/2}_0$ are the S--wave scattering lengths of
$\pi N$ scattering with isospin $I = 1/2$ and $I = 3/2$, $Q_{\pi^0n}=
28.040\,{\rm MeV}$ is the relative momentum of the $\pi^0 n$ pair at
relative momentum zero of the $\pi^-p$ pair, and $\Psi_{1s}(0) =
1/\sqrt{\pi a^3_B}$ is the wave function of pionic hydrogen in the
ground state at the origin $r = 0$\,\footnote{Here $a_B = 1/\alpha \mu
= 222.664\,{\rm fm}$ is the Bohr radius, where $\alpha = e^2 =
1/137.036$ is the fine--structure constant and $\mu =
m_{\pi^-}m_p/(m_{\pi^-} + m_p) = 121.497\,{\rm MeV}$ is the reduced
mass of the $\pi^-p$ pair calculated for $m_{\pi^-} = 139.570\,{\rm
MeV}$ and $m_p = 938.272\,{\rm MeV}$ \cite{KH02}.}. The imaginary part
of the amplitude $f^{\pi^-p}_0(0)$ in (\ref{label1.1}) is defined by
the inelastic channel $\pi^- + p \to \pi^0 + n$.

The DGBT formula (\ref{label1.1}) can be transcribed into an
equivalent form \cite{SD54} (see also \cite{TE88})
\begin{eqnarray}\label{label1.2}
\frac{\epsilon_{1s}}{E_{1s}} &=& + \frac{4}{a_B}\,A^{\pi^-p \to
\pi^-p}_S,\nonumber\\ \frac{\Gamma_{1s}}{E_{1s}} &=&-
\frac{8}{a_B}\,(A^{\pi^-p \to \pi^0n}_S)^2\,Q_{\pi^0n},
\end{eqnarray}
where $E_{1s} = - \alpha/2 a_B = - \alpha^2 \mu/2 = -\,3234.940\,{\rm
eV}$ is the binding energy of the ground state of pionic hydrogen and
\begin{eqnarray}\label{label1.3}
A^{\pi^-p \to \pi^-p}_S &=& \frac{1}{3}\,(2 a^{1/2}_0 +
a^{3/2}_0),\nonumber\\ A^{\pi^-p \to \pi^0n}_S &=&
\frac{\sqrt{2}}{3}\,(a^{3/2}_0 - a^{1/2}_0)
\end{eqnarray}
are the S--wave scattering lengths of $\pi^-p$ scattering.

The theoretical accuracy of the DGBT formula with respect to
next--to--leading order corrections caused by strong low--energy
interactions has been recently analysed in \cite{AI03}. As has been
shown the second order correction to the shift of the energy level of
the ground state relative to the first order makes up $(0.111\pm
0.006)\,\%$. In turn, the derivation of the energy level displacement
of the ground state of pionic hydrogen, carried out within a quantum
field theoretic, relativistic covariant and model--independent
approach \cite{AI03}, leads to the non--perturbative correction of
order of $1\%$. Hence, strong low--energy interactions cannot compete
with contributions of QCD isospin--breaking and electromagnetic
interactions \cite{JG02}. The predicted value of these corrections is
$\delta_{\epsilon} = (- 7.3\pm 2.9)\%$ \cite{JG02}\,\footnote{It is
likely that the correction obtained in \cite{AI03} is also included in
that calculated by Gasser {\it et al.}  \cite{JG02} within Chiral
Perturbation Theory (ChPT) \cite{JG99}. The analysis of this question
is in progress.}.

Experimentally \cite{PSI1,PSI2} the energy level displacement of the
ground state $(\epsilon_{1s}, \Gamma_{1s})$ can be obtained by
measuring the $np \to 1s$ transitions in pionic hydrogen for $n =
2,3,4$, where $n$ is the principle quantum number of the bound $np$
state of pionic hydrogen with the angular momentum $\ell = 1$. As a
result, the energy level displacements $(\epsilon_{np},\Gamma_{np})$
of the excited $np$ states turn out to be entangled into the
definition of $(\epsilon_{1s}, \Gamma_{1s})$.

Within a potential model approach the shift of the energy level of the
excited $ns$ state of pionic hydrogen has been calculated by Trueman
\cite{TT61} and Ericson, Loiseau and Wycech \cite{TE03}, who have also
given a systematic analysis of electromagnetic corrections, and of the
excited $n\ell$ state by Partensky and Ericson \cite{TE67}, Lambert
\cite{LA69} and Deloff \cite{DE76}. 

The main aim of this paper is (i) to derive the general formula for
the energy level displacement of the $n\ell$ excited state of pionic
hydrogen in terms of the partial--wave scattering lengths of $\pi N$
scattering within a quantum field theoretic, relativistic covariant
and model--independent approach developed in \cite{AI03} (see also
\cite{CD03}), (ii) to give an analytical expression and a numerical
value of the energy level displacement of the $np$ excited state of
pionic hydrogen, (iii) to calculate the second order correction to the
shift of the energy level of the ground state of pionic hydrogen
caused by the $ns \to 1s$ transitions and the $(\pi^-p)_{\rm Coul} \to
1s$ transitions of the $\pi^-p$ pair, coupled by the attractive Coulomb
field in the S--wave state with a continuous energy spectrum
\cite{TE03a}, and (iv) to compare our results with those obtained in
\cite{TT61}--\cite{TE88}.

As has been shown in \cite{AI03} (see also \cite{CD03}) the energy
level displacement of the ground state of pionic hydrogen can be
represented in terms of the momentum integrals
\begin{eqnarray}\label{label1.4}
-\epsilon_{1s} + i\,\frac{\Gamma_{1s}}{2} &=&
\frac{1}{4m_{\pi^-}m_p}\int
\frac{d^3k}{(2\pi)^3}\frac{d^3q}{(2\pi)^3}\,
\sqrt{\frac{m_{\pi^-}m_p}{E_{\pi^-}(k)E_p(k)}}\,
\sqrt{\frac{m_{\pi^-}m_p}{E_{\pi^-}(q)E_p(q)}}\nonumber\\
&&\times\,\Phi^{\dagger}_{1s}(\vec{k}\,)\,M(\pi^-(\vec{q}\,)
p(-\vec{q},\sigma_p) \to \pi^-(\vec{k}\,)
p(-\vec{k},\sigma_p))\,\Phi_{1s}(\vec{q}\,),
\end{eqnarray}
where $M(\pi^-(\vec{q}\,) p(-\vec{q},\sigma_p) \to \pi^-(\vec{k}\,)
p(-\vec{k},\sigma_p))$ is the amplitude of $\pi^-p$ scattering for
arbitrary relative momenta of the $\pi^-p$ pair, $E_{\pi^-} =
\sqrt{k^2 + m^2_{\pi^-}}$ and $E_p(k) = \sqrt{k^2 + m^2_p}$ are the
energies of the $\pi^-$--meson and the proton, $\sigma_p = \pm 1/2$ is
polatization of the proton, $\Phi_{1s}(\vec{k}\,)$ is the wave
function of the ground state of pionic hydrogen in the momentum
representation normalized by
\begin{eqnarray}\label{label1.5}
\int \frac{d^3k}{(2\pi)^3}\,|\Phi_{1s}(\vec{k}\,)|^2 = 1,
\end{eqnarray}
Near threshold the r.h.s. of (\ref{label1.4}) can be rewritten as
follows
\begin{eqnarray}\label{label1.6}
-\epsilon_{1s} + i\,\frac{\Gamma_{1s}}{2} &=& \frac{2\pi}{\mu}\int
\frac{d^3k}{(2\pi)^3}\frac{d^3q}{(2\pi)^3}\,
\sqrt{\frac{m_{\pi^-}m_p}{E_{\pi^-}(k)E_p(k)}}\,
\sqrt{\frac{m_{\pi^-}m_p}{E_{\pi^-}(q)E_p(q)}}\nonumber\\
&&\times\,\Phi^{\dagger}_{1s}(\vec{k}\,)\,f^{\pi^-p}_0(\sqrt{kq})\,
\Phi_{1s}(\vec{q}\,),
\end{eqnarray}
where we have set \cite{HP73}
\begin{eqnarray}\label{label1.7}
M(\pi^-(\vec{q}\,) p(-\vec{q},\sigma_p) \to \pi^-(\vec{k}\,)
p(-\vec{k},\sigma_p)) = 8\pi\,(m_{\pi^-} +
m_p)\,f^{\pi^-p}_0(\sqrt{kq}).
\end{eqnarray}
Due to the wave functions $\Phi^{\dagger}_{1s}(\vec{k}\,)$ and
$\Phi_{1s}(\vec{q}\,)$ the integrand of the momentum integrals in
(\ref{label1.4}) and (\ref{label1.6}) is concentrated around $k \sim q
\sim 1/a_B = 0.887\,{\rm MeV}$.  This justifies the application of the
low--energy limit $k,q \to 0$ to the calcuation of the amplitude of
$\pi^-p$ scattering \cite{AI03}. As a result the r.h.s. of
(\ref{label1.6}) reduces to the form of the DGBT formula with an
additional non--perturbative correction of order of 1$\%$, caused by
the smearing of the wave function of the ground state of pionic
hydrogen around the origin \cite{AI03}.

The paper is organized as follows.  In Section 2 we construct the wave
function of the excited $n\ell$ state of pionic hydrogen following the
prescription developed in \cite{AI03}. In Sections 3 and 4 following
\cite{AI03} we calculate the shift and the width of the energy level
of the excited $n\ell$ state of pionic hydrogen within a quantum field
theoretic, relativistic covariant and model--independent approach
developed in \cite{AI03}.  In Section 5 we calculate analytically and
give numerical estimate of the energy level displacement of the
excited $np$ state of pionic hydrogen. The shift of the energy level
of the $np$ state is found in analytical and numerical agreement with
the result obtained by Ericson and Weise within a potential model
approach \cite{TE88}. In Section 6 we calculate the contribution of
the $ns \to 1s$ transitions and the $(\pi^-p)_{\rm Coul} \to 1s$
transitions of the $\pi^-p$ pair, coupled by the attractive Coulomb
field in the S--wave state with a continuous energy spectrum, to the
shift of the energy level of the ground state of pionic hydrogen
induced by strong low--energy interactions.  We find that this
contribution relative to the DGBT shift makes up $0.076\,\%$. In the
Conclusion we discuss the obtained results. In the Appendix we
calculate the momentum integral defining the energy level displacement
of the $np$ state of pionic hydrogen.

\section{Wave function of $n\ell$ state of pionic hydrogen}
\setcounter{equation}{0}

The wave function of pionic hydrogen in the $1s$ state
has been defined as \cite{AI03}
\begin{eqnarray}\label{label2.1}
|A^{(1s)}_{\pi p}(\vec{P},\sigma_p)\rangle &=& \frac{1}{(2\pi)^3}\int
 \frac{d^3k_{\pi^-}}{\sqrt{2E_{\pi^-}(\vec{k}_{\pi^-})}}
 \frac{d^3k_p}{\sqrt{2E_p(\vec{k}_p)}} \delta^{(3)}(\vec{P} -
 \vec{k}_{\pi^-} - \vec{k}_p)\,\nonumber\\
 &&\times\,\sqrt{2E^{(1s)}_A(\vec{k}_{\pi^-} +\vec{k}_p) }\,
 \Phi_{1s}(\vec{k}_{\pi^-})|\pi^-(\vec{k}_{\pi^-})
 p(\vec{k}_p,\sigma_p)\rangle,
\end{eqnarray}
where $E^{(1s)}_A(\vec{P}\,) = \sqrt{{M^{(1s)}_A}^{\textstyle _2} +
\vec{P}^{\;2}}$ and $\vec{P}$ are the total energy and the momentum of
pionic hydrogen, $M^{(1s)}_A = m_p + m_{\pi^-} + E_{1s}$ is the mass
of pionic hydrogen in the $1s$ bound state, $\sigma_p = \pm 1/2$ is
the proton polarization; $\Phi_{1s}(\vec{k}_{\pi^-})$ is the wave
function of the $1s$ state of pionic hydrogen in the momentum
representation. It is normalized to unity (\ref{label1.5}).  The wave
function $|\pi^-(\vec{k}_{\pi^-}) p(\vec{k}_p,\sigma_p)\rangle$ we
define as \cite{AI03}
\begin{eqnarray}\label{label2.2}
|\pi^-(\vec{k}_{\pi^-})p(\vec{k}_p,\sigma_p)\rangle =
 c^{\dagger}_{\pi^-}(\vec{k}_{\pi^-})a^{\dagger}_p(\vec{k}_p,
 \sigma_p)|0\rangle,
\end{eqnarray}
where $c^{\dagger}_{\pi^-}(\vec{k}_{\pi^-})$ and
 $a^{\dagger}_p(\vec{k}_p, \sigma_p)$ are creation operators of the
 $\pi^-$ meson with  momentum $\vec{k}_{\pi^-}$ and the proton with
 momentum $\vec{k}_p$ and polarization $\sigma_p = \pm 1/2$. They
 satisfy standard relativistic covariant commutation and
 anti--commutation relations
\begin{eqnarray}\label{label2.3}
&&[c_{\pi^-}(\vec{k}\,'_{\pi^-}),
c^{\dagger}_{\pi^-}(\vec{k}_{\pi^-})] = (2\pi)^3\,
2E_{\pi^-}(\vec{k}_{\pi^-})\,\delta^{(3)}(\vec{k}\,'_{\pi^-} -
\vec{k}_{\pi^-}),\nonumber\\ &&[c_{\pi^-}(\vec{k}\,'_{\pi^-}),
c_{\pi^-}(\vec{k}_{\pi^-})] =
[c^{\dagger}_{\pi^-}(\vec{k}\,'_{\pi^-}),
c^{\dagger}_{\pi^-}(\vec{k}_{\pi^-})] = 0,\nonumber\\
&&\{a_p(\vec{k}\,'_p, \sigma\,'_p), a^{\dagger}_p(\vec{k}_p,
\sigma_p)\} = (2\pi)^3\, 2E_p(\vec{k}_p)\,\delta^{(3)}(\vec{k}\,'_p -
\vec{k}_p)\,\delta_{\sigma\,'_p\sigma_p},\nonumber\\
&&\{a_p(\vec{k}\,'_p, \sigma\,'_p), a_p(\vec{k}_p, \sigma_p)\} =
\{a^{\dagger}_p(\vec{k}\,'_p, \sigma\,'_p), a^{\dagger}_p(\vec{k}_p,
\sigma_p)\} = 0.
\end{eqnarray}
The wave function (\ref{label2.1}) is normalized by
\begin{eqnarray}\label{label2.4}
\langle A^{(1s)}_{\pi p}(\vec{P}\,',\sigma\,'_p)|A^{(1s)}_{\pi
p}(\vec{P},\sigma_p)\rangle =  (2\pi)^3\,
2E^{(1s)}_A(\vec{P}\,)\,\delta^{(3)}(\vec{P}\,' -
\vec{P}\,)\,\delta_{\sigma\,'_p\sigma_p}.
\end{eqnarray}
This is a relativistic covariant normalization of the wave function.
The wave function (\ref{label2.1}) has been applied to the derivation
of the energy level displacement of the ground state of pionic
hydrogen within a quantum field theoretic, relativistic covariant
and model--independent approach \cite{AI03}.

In analogy with the $1s$ state we define the wave function of the
$n\ell$ excited state of pionic hydrogen
\begin{eqnarray}\label{label2.5}
|A^{(n\ell m)}_{\pi p}(\vec{P},\sigma_p)\rangle &=&
 \frac{1}{(2\pi)^3}\int
 \frac{d^3k_{\pi^-}}{\sqrt{2E_{\pi^-}(\vec{k}_{\pi^-})}}
 \frac{d^3k_p}{\sqrt{2E_p(\vec{k}_p)}} \delta^{(3)}(\vec{P} -
 \vec{k}_{\pi^-} - \vec{k}_p)\,\nonumber\\
 &&\times\,\sqrt{2E^{(n\ell)}_A(\vec{k}_{\pi^-} +\vec{k}_p) }\,
 \Phi_{n\ell m}(\vec{k}_{\pi^-})|\pi^-(\vec{k}_{\pi^-})
 p(\vec{k}_p,\sigma_p)\rangle,
\end{eqnarray}
where $E^{(n\ell)}_A(\vec{P}\,) = \sqrt{{M^{(n\ell)}_A}^{\textstyle
_2} + \vec{P}^{\;2}}$ and $\vec{P}$ are the total energy and the
momentum of pionic hydrogen, $M^{(n\ell)}_A = m_p + m_{\pi^-} +
E_{n\ell}$ and $E_{n\ell}$ are the mass and the binding energy of
pionic hydrogen in the $n\ell$ bound state.  $\Phi_{n\ell
m}(\vec{k}\,)$ is the wave function of the excited $n\ell$ state in
the momentum representation, where $n$ is the {\it principle quantum
number}, $\ell$ is the angular momentum $\ell = 0,1, \ldots, n - 1$
and $m$ is the magnetic quantum number $m = 0, \pm 1, \ldots, \pm
\ell$.

The wave function of pionic hydrogen in the coordinate representation
$\Psi_{n\ell m}(\vec{r}\,)$ is equal to \cite{LL65}
\begin{eqnarray}\label{label2.6}
\Psi_{n\ell m}(\vec{r}\,) &=& R_{n\ell}(r)\,Y_{\ell
m}(\vartheta,\varphi) =\nonumber\\ &=& - \frac{2}{n^2}\,\sqrt{\frac{(n
- \ell - 1)!}{[(n + \ell)!]^3 a^3_B}}\,\Big(\frac{2 }{n }\frac{r}{
a_B}\Big)^{\ell}e^{\textstyle\,-r/n a_B}\,L^{2\ell + 1}_{n +
\ell}\Big(\frac{2 }{n }\frac{r}{ a_B}\Big)\,Y_{\ell
m}(\vartheta,\varphi),
\end{eqnarray}
where $L^{2\ell + 1}_{n + \ell}(\rho)$ are the generalized Laguerre
polynomials defined by \cite{LL65}
\begin{eqnarray}\label{label2.7}
L^{2\ell + 1}_{n + \ell}(\rho) = (-1)^{2\ell + 1}\,\frac{(n +
\ell)!}{(n - \ell - 1)!}\,\rho^{-(2\ell +
1)}\,e^{\textstyle\,\rho}\,\frac{d^{n - \ell - 1}}{d\rho^{n - \ell -
1}}(\rho^{n + \ell}e^{\textstyle\,-\rho}),
\end{eqnarray}
and $Y_{\ell m}(\vartheta,\varphi)$ are spherical harmonics normalized
by
\begin{eqnarray}\label{label2.8}
\int d\Omega\,Y^*_{\ell\,' m\,'}(\vartheta,\varphi)Y_{\ell
m}(\vartheta,\varphi) = \delta_{\ell\,'\ell}\,\delta_{m\,'m},
\end{eqnarray}
where $d\Omega = \sin\vartheta d\vartheta d\varphi$ is a solid angle.

In the momentum representation the wave function $\Phi_{n\ell
m}(\vec{k}\,)$ of pionic hydrogen is determined as
\begin{eqnarray}\label{label2.9}
\hspace{-0.3in}&&\Phi_{n\ell m}(\vec{k}\,) = \int d^3x\,\Psi_{n\ell
m}(\vec{r}\,)\,e^{\textstyle\,-i\vec{k}\cdot \vec{r}} =
\int^{\infty}_0dr r^2 R_{n\ell}(r)\nonumber\\ \hspace{-0.3in}&&\times
\int d\Omega\,\sum^{\infty}_{\ell\,' = 0}\sum^{\ell\,'}_{m\,' = -
\ell\,'}4\pi\, (-
i)^{\ell\,'}j_{\ell\,'}(kr)\,Y^*_{\ell\,'m\,'}(\vartheta,\varphi)
Y_{\ell\,'m\,'}(\vartheta_{\vec{k}},\varphi_{\vec{k}})Y_{\ell m
}(\vartheta,\varphi) =\nonumber\\\hspace{-0.3in}&&= (-
i)^{\ell}\,Y_{\ell m }(\vartheta_{\vec{k}},\varphi_{\vec{k}})\,4\pi
\int^{\infty}_0j_{\ell}(kr)R_{n\ell}(r)r^2 dr = (-
i)^{\ell}\,\sqrt{4\pi}\,\Phi_{n\ell}(k)Y_{\ell m
}\,(\vartheta_{\vec{k}},\varphi_{\vec{k}}),
\end{eqnarray}
where $j_{\ell}(kr)$ are spherical Bessel functions \cite{MA72},
$Y_{\ell m }(\vartheta_k,\varphi_k)$ and $\Phi_{n\ell}(k)$ are
spherical harmonics and radial wave functions in momentum space. The
radial wave functions $\Phi_{n\ell}(k)$ are defined as
\begin{eqnarray}\label{label2.10}
\Phi_{n\ell}(k) =\sqrt{4\pi}
\int^{\infty}_0j_{\ell}(kr)R_{n\ell}(r)r^2 dr.
\end{eqnarray}
Now we are able to proceed to calculating the energy level
displacement of the excited $n\ell$ state of pionic hydrogen.

\section{Shift of energy level of excited $n\ell$ state}
\setcounter{equation}{0}

The shift of the energy level $\epsilon_{n\ell}$ of the excited
$n\ell$ state of pionic we define as \cite{AI03}
\begin{eqnarray}\label{label3.1}
\epsilon_{n\ell} = - \frac{1}{2M^{(n\ell)}_A} \frac{1}{2\ell +
1}\sum^{\ell}_{m = - \ell}\langle A^{(n\ell m)}_{\pi
p}(0,\sigma_p)|{\cal L}_{\rm str}(0)|A^{(n\ell m)}_{\pi
p}(0,\sigma_p)\rangle,
\end{eqnarray}
where ${\cal L}_{\rm str}(x)$ is an effective total Lagrangian of
strong low--energy interactions. For the quantum field theoretic and
model--independent calculation of the shift of the energy level of the
$n\ell$ state of pionic hydrogen we will not specify ${\cal L}_{\rm
str}(x)$ in terms of interpolating fields of the coupled mesons and
baryons. We would like to emphasize that ${\cal L}_{\rm str}(x)$ is a
total {\it effective} Lagrangian accounting for all strong low--energy
interactions. In other words this {\it effective} Lagrangian defines
the $\mathbb{T}_{\rm str}$--matrix of strong low--energy interactions
\begin{eqnarray}\label{label3.2}
\mathbb{T}_{\rm str} = \int d^4x\,{\cal L}_{\rm str}(x)
\end{eqnarray}
obeying the unitary condition \cite{SS61,HP73}
\begin{eqnarray}\label{label3.3}
\mathbb{T}_{\rm str} - \mathbb{T}^{\dagger}_{\rm str} =
i\mathbb{T}^{\dagger}_{\rm str}\mathbb{T}_{\rm str}.
\end{eqnarray}
This means that the matrix element of the {\it effective} Lagrangian
${\cal L}_{\rm str}(0)$ between the states $|\pi^-p\rangle$ defines a
physical amplitude of $\pi^- p$ scattering \cite{AI03}
\begin{eqnarray}\label{label3.4}
\langle \pi^- p|{\cal L}_{\rm str}(0)|\pi^- p\rangle = \frac{1}{3}\,(2
T^{1/2} + T^{3/2})
\end{eqnarray}
where $T^{1/2}$ and $T^{3/2}$ are the amplitudes of $\pi N$ scattering
with isotopic spin $I = 1/2$ and $I = 3/2$, respectively.

According to \cite{AI03} the shift of the energy level of the $n\ell$
state, expressed in terms of the matrix element of the effective
Lagrangian ${\cal L}_{\rm str}(0)$, reads
\begin{eqnarray}\label{label3.5}
\hspace{-0.3in}\epsilon_{n\ell} &=& - \frac{1}{2\ell +
1}\sum^{\ell}_{m = -\ell}\int \frac{d^3k}{(2\pi)^3}
\frac{\Phi^{\dagger}_{n \ell m}(\vec{k}\,)}{\sqrt{2
E_{\pi^-}(\vec{k}\,)2 E_p(\vec{k}\,)}}\int \frac{d^3q}{(2\pi)^3}
\frac{\Phi_{n\ell m}(\vec{q}\,)}{\sqrt{2 E_{\pi^-}(\vec{q}\,)2
E_p(\vec{q}\,)}}\nonumber\\
\hspace{-0.3in}&&\times \langle
\pi^-(\vec{k}\,)p(-\vec{k},\sigma_p)|{\cal L}_{\rm
str}(0)|\pi^-(\vec{q}\,)p(-\vec{q},\sigma_p)\rangle
\end{eqnarray}
It is convenient to redefine the r.h.s. of (\ref{label3.5}) as follows
\begin{eqnarray}\label{label3.6}
\hspace{-0.5in}&& \epsilon_{n\ell}= - \frac{1}{2\ell +
1}\sum^{\ell}_{m = -\ell}\int \frac{d^3k}{(2\pi)^3}
\frac{\Phi^{\dagger}_{n \ell}(k)}{\sqrt{2 E_{\pi^-}(k)2 E_p(k)}}\int
\frac{d^3q}{(2\pi)^3} \frac{\Phi_{n\ell}(q)}{\sqrt{2 E_{\pi^-}(q)2
E_p(q)}}\nonumber\\
\hspace{-0.5in}&&\times \int\!\!\!\int
\frac{d\Omega_{\vec{k}}}{\sqrt{4\pi}}\frac{d\Omega_{\vec{q}}}{\sqrt{4\pi}}\,Y^*_{\ell
m}(\vartheta_{\vec{k}},\varphi_{\vec{k}})\,\langle
\pi^-(\vec{k}\,)p(-\vec{k},\sigma_p)|{\cal L}_{\rm
str}(0)|\pi^-(\vec{q}\,)p(-\vec{q},\sigma_p)\rangle\,Y_{\ell
m}(\vartheta_{\vec{q}},\varphi_{\vec{q}}).
\end{eqnarray}
Since there is no spin--flip in the low--energy transition $\pi^- + p
\to \pi^- + p$ the amplitude of $\pi^-p$ scattering is determined by
\cite{SG67,MN79}
\begin{eqnarray}\label{label3.7}
\hspace{-0.3in}&&\langle \pi^-(\vec{k}\,)p(-\vec{k},\sigma_p)|{\cal
L}_{\rm str}(0)|\pi^-(\vec{q}\,)p(-\vec{q},\sigma_p)\rangle =
8\pi\,\sqrt{s}\sum^{\infty}_{\ell\,' = 0}[(\ell\,' +
1)\,f_{\ell\,'+}(\sqrt{kq}) + \ell\,'\,f_{\ell\,'-}(\sqrt{kq})]
\nonumber\\
\hspace{-0.3in}&&\times\,P_{\ell\,'}(\cos\vartheta)=
8\pi\,\sqrt{s}\sum^{\infty}_{\ell\,' = 0}[(\ell\,' +
1)\,f_{\ell\,'+}(\sqrt{kq}) + \ell\,'\,f_{\ell\,'-}(\sqrt{kq})]\,
\sum^{\ell\,'}_{m\,' = -\ell\,'}\frac{4\pi}{2 \ell\,' + 1}Y^*_{\ell\,'
m\,'}(\vartheta_{\vec{q}},\varphi_{\vec{q}})\nonumber\\
\hspace{-0.3in}&&\times\,Y_{\ell\,'
m\,'}(\vartheta_{\vec{k}},\varphi_{\vec{k}}),
\end{eqnarray}
where $\sqrt{s}$ is the total energy in the $s$--channel of $\pi^-p$
scattering, $P_{\ell\,'}(\cos\vartheta)$ are Legendre polynomials
\cite{MA72} and $\vartheta$ is the angle between the relative momenta
$\vec{k}$ and $\vec{q}$. The amplitudes $f_{\ell\,'+}(\sqrt{kq})$ and
$f_{\ell\,'-}(\sqrt{kq})$ describe $\pi^-p$ scattering in the states
with a total momentum $J = \ell\,' + 1/2$ and $J = \ell\,' -
1/2$. They are defined by the phase shifts \cite{SG67,MN79}
\begin{eqnarray}\label{label3.8}
f_{\ell\,'+}(\sqrt{kq}) &=& \frac{\displaystyle
e^{\textstyle\,+i\delta_{\ell\,'+}(\sqrt{kq})}}{\sqrt{kq}}\,
\sin\delta_{\ell\,'+}(\sqrt{kq}),\nonumber\\ f_{\ell\,'-}(\sqrt{kq}) &=&
\frac{\displaystyle
e^{\textstyle\,+i\delta_{\ell\,'-}(\sqrt{kq})}}{\sqrt{kq}}\,
\sin\delta_{\ell\,'-}(\sqrt{kq}).
\end{eqnarray}
Near threshold $k, q \to 0$ the amplitudes $f_{\ell\,'+}(\sqrt{kq})$
and $f_{\ell\,'-}(\sqrt{kq})$ are defined by the $\ell\,'$--wave
scattering lengths of $\pi^- p$ scattering \cite{MN79}
\begin{eqnarray}\label{label3.9}
f_{\ell\,'+}(\sqrt{kq}) &=&
a^{\pi^-p\to\pi^-p}_{\ell\,'+}(kq)^{\,\ell\,'},\nonumber\\
f_{\ell\,'-}(\sqrt{kq}) &=&
a^{\pi^-p\to\pi^-p}_{\ell\,'-}(kq)^{\,\ell\,'}.
\end{eqnarray}
Substituting (\ref{label3.7}) and (\ref{label3.9}) in (\ref{label3.6})
and integrating over the solid angles we get the shift of the energy
level of the excited $n\ell$ state
\begin{eqnarray}\label{label3.10}
\hspace{-0.3in}\epsilon_{n\ell} = - \frac{2\pi}{\mu}\,\frac{(\ell +
1)\,a^{\pi^-p\to\pi^-p}_{\ell+} +
\ell\,a^{\pi^-p\to\pi^-p}_{\ell-}}{2\ell + 1}\,\Bigg|\int
\frac{d^3k}{(2\pi)^3} \sqrt{\frac{m_{\pi^-} m_p}{
E_{\pi^-}(k)E_p(k)}}\,k^{\,\ell}\,\Phi_{n \ell}(k)\Bigg|^2.
\end{eqnarray}
This is a generalization of the DGBT formula to any excited $n\ell$
state of pionic hydrogen.

The $\ell$--wave scattering lengths $a^{\pi^-p\to\pi^-p}_{\ell \pm}$
are related to the $\ell$--wave scattering lengths $a^I_{\ell \pm}$
for $I = 1/2$ and $I = 3/2$ as
\begin{eqnarray}\label{label3.11}
a^{\pi^-p\to\pi^-p}_{\ell \pm} = \frac{1}{3}\,(2\,a^{1/2}_{\ell \pm} +
a^{3/2}_{\ell \pm}).
\end{eqnarray}
For the ground state $n = 1$ and $\ell = 0$ the expression
(\ref{label3.10}) with (\ref{label3.11}) reduces to the DGBT formula
(\ref{label1.1}) \cite{AI03}.

\section{Width of energy level of excited $n\ell$ state}
\setcounter{equation}{0}

According to \cite{AI03} the width $\Gamma_{n\ell}$ of
the energy level of the excited $n\ell$ state is defined by
\begin{eqnarray}\label{label4.1}
&&\Gamma_{n\ell} = \frac{1}{2\ell + 1}\sum^{\ell}_{m = -\ell}\int
\frac{d^3k}{(2\pi)^3}\frac{\Phi^{\dagger}_{n\ell
m}(\vec{k}\,)}{\sqrt{2 E_{\pi^-}(\vec{k}\,)2 E_p(\vec{k}\,)}}\int
\frac{d^3q}{(2\pi)^3}\frac{\Phi_{n\ell m}(\vec{q}\,)}{\sqrt{2
E_{\pi^-}(\vec{q}\,)2 E_p(\vec{q}\,)}}\nonumber\\ &&\times
\sum_{\lambda_n = \pm 1/2}\int \frac{d^3Q}{(2\pi)^3 2
E_{\pi^0}(\vec{Q})2 E_n(\vec{Q})}\,2\pi\,\delta(E_{\pi^0}(\vec{Q}) +
E_n(\vec{Q})- m_{\pi^-} - m_p + E_{n\ell})\nonumber\\ &&\times\langle
\pi^-(\vec{k}\,)p(-\vec{k},\sigma)|{\cal L}_{\rm
str}(0)|\pi^0(\vec{Q})n(- \vec{Q},\lambda_n)\rangle \langle n(-
\vec{Q}, \lambda_n) \pi^0(\vec{Q})|{\cal L}_{\rm
str}(0)|\pi^-(\vec{q}\,)p(-\vec{q},\sigma)\rangle.  \nonumber\\ &&
\end{eqnarray}
For the subsequent calculation it is convenient to rewrite the
r.h.s. of (\ref{label4.1}) as follows
\begin{eqnarray}\label{label4.2}
&&\Gamma_{n\ell} = \frac{1}{16\pi}\,\frac{1}{2\ell + 1}\sum^{\ell}_{m
= -\ell}\int \frac{d^3k}{(2\pi)^3}\frac{\Phi^{\dagger}_{n\ell}(k)}
{\sqrt{E_{\pi^-}(k) E_p(k)}}\int
\frac{d^3q}{(2\pi)^3}\frac{\Phi_{n\ell}(q)}{\sqrt{ E_{\pi^-}(q)
E_p(q)}}\nonumber\\ &&\times \sum_{\lambda_n = \pm 1/2}\int^{\infty}_0
\frac{dQ\,Q^2}{ E_{\pi^0}(Q)E_n(Q)}\,\delta(E_{\pi^0}(Q) + E_n(Q)-
m_{\pi^-} - m_p - E_{n\ell})\nonumber\\ &&\times\int
\frac{d\Omega_{\vec{Q}}}{4\pi}\int
\frac{d\Omega_{\vec{k}}}{\sqrt{4\pi}}\int
\frac{d\Omega_{\vec{q}}}{\sqrt{4\pi}}\, Y^*_{\ell
m}(\vartheta_{\vec{k}},\varphi_{\vec{k}})\,\langle
\pi^-(\vec{k}\,)p(-\vec{k},\sigma)|{\cal L}_{\rm
str}(0)|\pi^0(\vec{Q})n(- \vec{Q},\lambda_n)\rangle\nonumber\\
&&\times \langle n(- \vec{Q}, \lambda_n) \pi^0(\vec{Q})|{\cal L}_{\rm
str}(0)|\pi^-(\vec{q}\,)p(-\vec{q},\sigma)\rangle\, Y_{\ell
m}(\vartheta_{\vec{q}},\varphi_{\vec{q}}).
\end{eqnarray}
The matrix element $\langle \pi^-(\vec{k}\,)p(-\vec{k},\sigma)|{\cal
L}_{\rm str}(0)|\pi^0(\vec{Q})n(- \vec{Q},\lambda_n)\rangle$ we define
as 
\begin{eqnarray}\label{label4.3}
\hspace{-0.3in}&&\langle \pi^-(\vec{k}\,)p(-\vec{k},\sigma)|{\cal
L}_{\rm str}(0)|\pi^0(\vec{Q})n(- \vec{Q},\lambda_n)\rangle =
8\pi\,\sqrt{s}\sum^{\infty}_{\ell\,' = 0}[(\ell\,' +
1)\,f_{\ell\,'+}(\sqrt{kQ}) + \ell\,'\nonumber\\
&&\times\,f_{\ell\,'-}(\sqrt{kQ})] \,P_{\ell\,'}(\cos\vartheta)=
8\pi\,\sqrt{s}\sum^{\infty}_{\ell\,' = 0}[(\ell\,' +
1)\,f_{\ell\,'+}(\sqrt{kQ}) +
\ell\,'\,f_{\ell\,'-}(\sqrt{kQ})]\nonumber\\
&&\times\,\,\sum^{\ell\,'}_{m\,' = -\ell\,'}\frac{4\pi}{2 \ell\,' +
1}Y^*_{\ell\,'
m\,'}(\vartheta_{\vec{Q}},\varphi_{\vec{Q}})\,Y_{\ell\,'
m\,'}(\vartheta_{\vec{k}},\varphi_{\vec{k}}),
\end{eqnarray}
Near threshold the amplitudes $f_{\ell\,'+}(\sqrt{kQ})$ and
$f_{\ell\,'-}(\sqrt{kQ})$ are defined by the $\ell\,'$--wave scattering
lengths of the $\pi^- p \to \pi^0 n$ scattering \cite{MN79}
\begin{eqnarray}\label{label4.4}
f_{\ell\,'+}(\sqrt{kQ}) &=&
a^{\pi^-p\to\pi^0n}_{\ell\,'+}(kQ)^{\,\ell\,'},\nonumber\\
f_{\ell\,'-}(\sqrt{kq}) &=& a^{\pi^-p\to\pi^0n}_{\ell\,'-}
(kQ)^{\,\ell\,'}.
\end{eqnarray}
The $\ell\,'$--wave scattering lengths $a^{\pi^-p\to\pi^0n}_{\ell\,'
\pm}$ are related to the $\ell\,'$--wave scattering lengths
$a^I_{\ell\,' \pm}$ for $I = 1/2$ and $I = 3/2$ as
\begin{eqnarray}\label{label4.5}
a^{\pi^-p\to\pi^0n}_{\ell\,' \pm} =
\frac{\sqrt{2}}{3}\,(a^{3/2}_{\ell\,' \pm} - a^{1/2}_{\ell\,' \pm}).
\end{eqnarray}
Substituting (\ref{label4.3}) and (\ref{label4.4}) in (\ref{label4.2})
and integrating over the solid angles and the phase volume we arrive
at the expression
\begin{eqnarray}\label{label4.6}
\Gamma_{n\ell} &=& \frac{4\pi}{\mu}\,\Big[\frac{(\ell +
1)\,a^{\pi^-p\to\pi^0n}_{\ell+} +
\ell\,a^{\pi^-p\to\pi^0n}_{\ell-}}{2\ell + 1}\Big]^2\,Q^{\,2\ell +
1}_{n\ell}\nonumber\\ &&\times\,\Bigg|\int
\frac{d^3k}{(2\pi)^3}\frac{m_{\pi^-}m_p} {\sqrt{E_{\pi^-}(k)
E_p(k)}}\,k^{\,\ell}\,\Phi_{n\ell}(k)\Bigg|^2,
\end{eqnarray}
where $Q_{n\ell}$ is a relative momentum of the $\pi^0n$ pair
\begin{eqnarray}\label{label4.7}
Q_{n\ell} = \sqrt{\frac{2\,m_{\pi^0}\,m_n}{m_{\pi^0} +
m_n}\,(m_{\pi^-} + m_p - m_{\pi^0} - m_n + E_{n\ell})},
\end{eqnarray}
where $m_{\pi^0} = 134.977\,{\rm MeV}$ and $m_n = 939.565\,{\rm MeV}$
\cite{KH02}.  

This is a generalization of the DGBT formula to any excited $n\ell$
state of pionic hydrogen. For the ground state $n = 1$ and $\ell = 0$
we arrive at the DGBT formula (\ref{label1.1}).

\section{Energy level displacement of excited $np$ state}
\setcounter{equation}{0}

The analysis of experimental data obtained by the PSI
Collaboration demands the knowledge of the energy level displacements
of the excited $np$ states. For $\ell = 1$ from the formulas
(\ref{label3.10}) and (\ref{label4.6}) one gets
\begin{eqnarray}\label{label5.1}
\hspace{-0.5in}\epsilon_{np} &=& -
\frac{2\pi}{\mu}\,\frac{1}{3}\,(2\,a^{\pi^-p\to\pi^-p}_{P+} +
a^{\pi^-p\to\pi^-p}_{P-})\,\Bigg|\int \frac{d^3k}{(2\pi)^3}
\sqrt{\frac{m_{\pi^-} m_p}{ E_{\pi^-}(k)E_p(k)}}\,k\,\Phi_{n
p}(k)\Bigg|^2 = \nonumber\\ \hspace{-0.5in}&=& -
\frac{2\pi}{9}\,\frac{1}{\mu}\,\Big[2(2 a^{1/2}_{P+} + a^{1/2}_{P-}) +
(2 a^{3/2}_{P+} + a^{3/2}_{P-})\Big]\,\Bigg|\int \frac{d^3k}{(2\pi)^3}
\sqrt{\frac{m_{\pi^-} m_p}{ E_{\pi^-}(k)E_p(k)}}\,k\,\Phi_{n
p}(k)\Bigg|^2,\nonumber\\ \Gamma_{np} &=&
\frac{4\pi}{9}\,\frac{Q^3_{np}}{\mu}\,(2\,a^{\pi^-p\to\pi^0n}_{P+} +
a^{\pi^-p\to\pi^0n}_{P-})^2\,\Bigg|\int \frac{d^3k}{(2\pi)^3}
\sqrt{\frac{m_{\pi^-} m_p}{ E_{\pi^-}(k)E_p(k)}}\,k\,\Phi_{n
p}(k)\Bigg|^2 = \nonumber\\
\hspace{-0.5in}&=& \frac{8\pi}{81}\,\frac{Q^3_{np}}{\mu}\,\Big[(2
a^{3/2}_{P+} + a^{3/2}_{P-})- (2 a^{1/2}_{P+} +
a^{1/2}_{P-})\Big]^2\,\Bigg|\int \frac{d^3k}{(2\pi)^3}
\sqrt{\frac{m_{\pi^-} m_p}{ E_{\pi^-}(k)E_p(k)}}\,k\,\Phi_{n
p}(k)\Bigg|^2,\nonumber\\
\hspace{-0.5in}&&
\end{eqnarray}
where $a^I_{P+}$ and $a^I_{P-}$ are the P--wave scattering lengths of
 $\pi N$ scattering with isospin $I$ and total momentum $J = 3/2$ and
 $J = 1/2$, respectively, \cite{SG67,MN79}.

The integral over $k$ is calculated in the Appendix. The result reads
\begin{eqnarray}\label{label5.2}
&&\int \frac{d^3k}{(2\pi)^3} \sqrt{\frac{m_{\pi^-} m_p}{
E_{\pi^-}(k)E_p(k)}}\,k\,\Phi_{n p}(k) = \sqrt{\frac{n^2 - 1}{\pi
n^5a^5_B}}.
\end{eqnarray}
Substituting (\ref{label5.2}) in (\ref{label5.1}) we obtain the shift
and the width of the energy level of the excited $np$ state. They read
\begin{eqnarray}\label{label5.3}
\epsilon_{np} &=& - \,\frac{2}{9}\,\frac{\alpha^5}{n^3}\,\Big(1 -
\frac{1}{n^2}\Big)\,\Big(\frac{m_{\pi^-}m_p}{m_{\pi^-} +
m_p}\Big)^4\,\Big[2(2 a^{1/2}_{P+} + a^{1/2}_{P-}) + (2 a^{3/2}_{P+} +
a^{3/2}_{P-})\Big],\nonumber\\ \Gamma_{np}
&=&~~\,\frac{8}{81}\,\frac{\alpha^5}{n^3}\,\Big(1 -
\frac{1}{n^2}\Big)\,\Big(\frac{m_{\pi^-}m_p}{m_{\pi^-} +
m_p}\Big)^4\,\Big[(2 a^{3/2}_{P+} + a^{3/2}_{P-})- (2 a^{1/2}_{P+} +
a^{1/2}_{P-})\Big]^2\nonumber\\
&&\times\,\Big[\frac{2\,m_{\pi^0}\,m_n}{m_{\pi^0} + m_n}\,(m_{\pi^-} +
m_p - m_{\pi^0} - m_n + E_{np})\Big]^{3/2}.
\end{eqnarray}
The shift and the width of the energy level of the $np$ excited state
(\ref{label5.3}) can be rewritten in the equivalent form
\begin{eqnarray}\label{label5.4}
\frac{\epsilon_{np}}{E_{np}} &=& +\,\frac{4}{n}\,\Big(1 -
\frac{1}{n^2}\Big)\,\frac{A^{\pi^-p \to \pi^-p}_P}{a^3_B},\nonumber\\
\frac{\Gamma_{np}}{E_{np}} &=&-\, \frac{8}{n}\,\Big(1 -
\frac{1}{n^2}\Big)\,\frac{(A^{\pi^-p \to
\pi^0n}_P)^2}{a^3_B}\,Q^3_{np},
\end{eqnarray}
where $E_{np} = - \alpha/2a_B n^2$ is the binding energy of the
excited $np$ state and
\begin{eqnarray}\label{label5.5}
A^{\pi^-p \to \pi^-p}_P &=& \frac{1}{9}\;[2(2 a^{1/2}_{P+} +
a^{1/2}_{P-}) + (2 a^{3/2}_{P+} + a^{3/2}_{P-})],\nonumber\\ A^{\pi^-p
\to \pi^0n}_P &=& \frac{\sqrt{2}}{9}\;[(2 a^{3/2}_{P+} +
a^{3/2}_{P-})- (2 a^{1/2}_{P+} + a^{1/2}_{P-})]
\end{eqnarray}
are the P--wave scattering lengths of $\pi^-p$ scattering \cite{TE88}.
The ratio $\epsilon_{np}/E_{np}$ in (\ref{label5.4}) is in analytical
agreement with the result obtained by Ericson and Weise (see
\cite{TE88} Eq.(6.29)).

In order to estimate the values of the shift and width of the excited
$np$ state we use the experimental data on the P--wave scattering
lengths compiled in Table 5.3 of Ref.\cite{MN79} (H\"ohler 78, input:
Karlsruhe--Helsinki analysis 78):
\begin{eqnarray}\label{label5.6}
&&a^{1/2}_{P-} = (-0.082 \pm 0.002)\,m^{-3}_{\pi^-}\quad,\quad
a^{1/2}_{P+} = (-0.032 \pm 0.001)\,m^{-3}_{\pi^-},\nonumber\\
&&a^{3/2}_{P-} = (-0.044 \pm 0.001)\,m^{-3}_{\pi^-}\quad,\quad
a^{3/2}_{P+} = (+ 0.215 \pm 0.003)\,m^{-3}_{\pi^-},\nonumber\\ &&2(2
a^{1/2}_{P+} + a^{1/2}_{P-}) + (2 a^{3/2}_{P+} + a^{3/2}_{P-}) =
(0.094\pm 0.008)\,m^{-3}_{\pi^-},\nonumber\\ &&~\,(2 a^{3/2}_{P+} +
a^{3/2}_{P-})- (2 a^{1/2}_{P+} + a^{1/2}_{P-}) = (0.532 \pm
0.007)\,m^{-3}_{\pi^-}.
\end{eqnarray}
This yields
\begin{eqnarray}\label{label5.7}
\epsilon_{np} &=&-\frac{1}{~n^3}\,\Big(1 - \frac{1}{n^2}\Big)\,\times
(3.47 \pm 0.30)\times 10^{-5}\,{\rm eV},\nonumber\\ \Gamma_{np} &=&
~~\frac{1}{~n^3}\,\Big(1 - \frac{1}{n^2}\Big)\,\times(3.71 \pm
0.10)\times 10^{-7}\,{\rm eV}.
\end{eqnarray}
Thus, the values of the energy level displacements of the excited $np$
states of pionic hydrogen are much smaller than $ 10^{-5}\,{\rm
eV}$\,\footnote{For pionium, the bound $\pi^-\pi^+$ state, the shift
of the energy level of the $np$ state has been recently calculated by
Julia Schweizer \cite{JS03} within Chiral Perturbation Theory
\cite{JG99}.}.The experimental value of the energy level displacement
of the ground state of pionic hydrogen is equal to \cite{PSI2}
\begin{eqnarray}\label{label5.8}
\epsilon^{\exp}_{1s} &=& -\,7.108\pm 0.036\,{\rm eV},\nonumber\\
\Gamma^{\exp}_{1s} &=& ~~~0.868\pm 0.054\,{\rm eV}.
\end{eqnarray}
with an accuracy about $0.5\%$ and 6.2$\%$ for the shift and the
width, respectively.

The level of accuracy in a new set of experiments is about $0.2\,\%$,
i.e. $(\Delta^{\exp}_{\rm shift} = \pm 0.014\,{\rm eV})$, for the
energy level shift and $1\%$, i.e. $(\Delta^{\exp}_{\rm width} = \pm
0.009\,{\rm eV)}$, for the energy level width \cite{PSI1}. Hence,
according to (\ref{label5.7}) the contributions of the energy level
displacements of the excited $np$ states to the transitions $np \to
1s$ with $n = 2,3,4$ can be neglected, since $|\Delta^{\exp}_{\rm
shift}| \gg |\epsilon_{np}|$ and $|\Delta^{\exp}_{\rm width}| \gg
\Gamma_{np}$.

\section{Energy shift of ground state caused by 
$(\pi^-p)_{ns} \to (\pi^-p)_{1s}$ and $(\pi^-p)_{\rm Coul} \to
(\pi^-p)_{1s}$ transitions} \setcounter{equation}{0}

In this section we calculate the shift of the energy level of the
ground state of pionic hydrogen $\delta \epsilon_{1s}$ caused by the
$(\pi^-p)_{ns} \to (\pi^-p)_{1s}$ and $(\pi^-p)_{\rm Coul} \to
(\pi^-p)_{1s}$ transitions induced by strong low--energy interactions,
where $(\pi^-p)_{ns}$ is a bound $ns$ state of the $\pi^-p$ pair and
$(\pi^-p)_{\rm Coul}$ is the $\pi^-p$ pair, coupled by the attractive
Coulomb field in the S--state with a continuous energy spectrum
\cite{TE03a}. According to \cite{AI03} the correction $\delta
\epsilon_{1s}$ reads
\begin{eqnarray}\label{label6.1}
\delta \epsilon_{1s} = - \frac{1}{~2 M^{(1s)}_A}\,\frac{i}{2}\int
d^4x\,\langle A^{(1s)}_{\pi p}(\vec{P},\sigma_p)|{\rm T}({\cal L}_{\rm
str}(x){\cal L}_{\rm str}(0))|A^{(1s)}_{\pi
p}(\vec{P},\sigma_p)\rangle\Big|_{\vec{P} = 0}.
\end{eqnarray}
First, we consider the contribution of the discrete spectrum. For this
aim we use a unit operator which we define as
\begin{eqnarray}\label{label6.2}
\hat{1} = \sum_{\alpha_p = \pm 1/2}\sum^{\infty}_{n = 1}\sum^{n -
1}_{\ell = 0}\sum^{\ell}_{m = - \ell}\frac{1}{(2\pi)^3}\int
\frac{d^3Q}{2 E^{(n\ell)}_A(\vec{Q}\,)}|A^{(n\ell m)}_{\pi p}(\vec{Q},
\alpha_p)\rangle \langle A^{(n\ell m)}_{\pi p}(\vec{Q}, \alpha_p)|.
\end{eqnarray}
Following \cite{AI03} and using a unit operator (\ref{label6.2}) for
the description of the intermediate states in (\ref{label6.1}), and
integrating over angular degrees of freedom we get
\begin{eqnarray}\label{label6.3}
\hspace{-0.3in}&&\delta \epsilon^{(\rm ds)}_{1s} = \sum_{\alpha_p =
\pm 1/2}{\cal P}\int \frac{d^3k}{(2\pi)^3}\,\frac{\Phi^*_{1s}(k)
}{\sqrt{2 E_{\pi^-}(k) 2 E_p(k)}}\int
\frac{d^3Q}{(2\pi)^3}\,\frac{\Phi_{1s}(Q) }{\sqrt{2 E_{\pi^-}(Q) 2
E_p(Q)}}\nonumber\\
\hspace{-0.3in}&&\times\,\frac{ \langle
\pi^-(\vec{k}\,)p(-\vec{k},\sigma_p)|{\cal L}_{\rm
str}(0)|p(-\vec{Q},\sigma_p)\pi^-(\vec{Q}\,)\rangle\langle
\pi^-(\vec{P}\,)p(-\vec{P},\sigma_p)|{\cal L}_{\rm
str}(0)|p(-\vec{q},\sigma_p)\pi^-(\vec{q}\,)
\rangle}{E_{\pi^-}(k) + E_p(k) -
E_{\pi^-}(Q) - E_p(Q)}\nonumber\\
\hspace{-0.3in}&&\times\,\int
\frac{d^3P}{(2\pi)^3}\,\frac{\Phi^*_{1s}(P) }{\sqrt{2 E_{\pi^-}(P) 2
E_p(P)}}\int \frac{d^3q}{(2\pi)^3}\,\frac{\Phi_{1s}(q) }{\sqrt{2
E_{\pi^-}(q) 2 E_p(q)}}\nonumber\\
\hspace{-0.3in}&&+ \sum_{\alpha_p = \pm 1/2}\sum^{\infty}_{n = 2}{\cal
P}\int \frac{d^3k}{(2\pi)^3}\,\frac{\Phi^*_{1s}(k) }{\sqrt{2
E_{\pi^-}(k) 2 E_p(k)}}\int \frac{d^3Q}{(2\pi)^3}\,\frac{\Phi_{ns}(Q)
}{\sqrt{2 E_{\pi^-}(Q) 2 E_p(Q)}}\nonumber\\
\hspace{-0.3in}&&\times\,\frac{ \langle
\pi^-(\vec{k}\,)p(-\vec{k},\sigma_p)|{\cal L}_{\rm
str}(0)|p(-\vec{Q},\sigma_p)\pi^-(\vec{Q}\,)\rangle\langle
\pi^-(\vec{P}\,)p(-\vec{P},\sigma_p)|{\cal L}_{\rm
str}(0)|p(-\vec{q},\sigma_p)\pi^-(\vec{q}\,)
\rangle}{E_{\pi^-}(k) + E_p(k) + E_{1s} -
E_{\pi^-}(Q) - E_p(Q) - E_{ns}}\nonumber\\
\hspace{-0.3in}&&\times\,\int
\frac{d^3P}{(2\pi)^3}\,\frac{\Phi^*_{ns}(P) }{\sqrt{2 E_{\pi^-}(P) 2
E_p(P)}}\int \frac{d^3q}{(2\pi)^3}\,\frac{\Phi_{1s}(q) }{\sqrt{2
E_{\pi^-}(q) 2 E_p(q)}},
\end{eqnarray}
where the abbreviation (ds) means the {\it discrete spectrum} and
${\cal P}$ stands for the calculation of the principle value of the
integral.

Due to the wave functions of pionic hydrogen the integrands in the
r.h.s. of (\ref{label6.3}) can be taken at the low--energy limit
\cite{AI03}. This yields
\begin{eqnarray}\label{label6.4}
\hspace{-0.5in}\delta \epsilon^{\rm (ds)}_{1s} &=&
\frac{8\pi^2}{9}\,\frac{1}{\mu}\,(2a^{1/2}_0 +
a^{3/2}_0)^2\,|\Psi_{1s}(0)|^2\,{\cal P}\int
\frac{d^3k}{(2\pi)^3}\frac{d^3Q}{(2\pi)^3}\,\frac{\Phi^*_{1s}(k)
\Phi_{1s}(Q)}{k^2 - Q^2}\nonumber\\
\hspace{-0.5in}&&+ \frac{4\pi^2}{9}\,\frac{1}{\mu^2}\,(2a^{1/2}_0 +
a^{3/2}_0)^2\,\sum^{\infty}_{n =
2}\frac{|\Psi^*_{ns}(0)\Psi_{1s}(0)|^2}{E_{1s} - E_{ns}}.
\end{eqnarray}
Since in the momentum representation the wave function of the ground
state of pionic hydrogen can be taken real, $\Phi^*_{1s}(k) =
\Phi_{1s}(k)$, the integrand of the first term in the r.h.s. of
(\ref{label6.4}) is antisymmetric under the change of variables $k
\longleftrightarrow Q$. Hence, the integral over $\vec{k}$ and
$\vec{Q}$ should vanish.

Thus, the shift of the energy level of the ground state, caused by the
$ns \to 1s$ transitions, is defined by
\begin{eqnarray}\label{label6.5}
\hspace{-0.5in}\delta \epsilon^{\rm (ds)}_{1s} =
\frac{4\pi^2}{9}\,\frac{1}{\mu^2}\,(2a^{1/2}_0 +
a^{3/2}_0)^2\,\sum^{\infty}_{n =
2}\frac{|\Psi^*_{ns}(0)\Psi_{1s}(0)|^2}{E_{1s} - E_{ns}}.
\end{eqnarray}
Setting $\Psi_{1s}(0) = 1/\sqrt{\pi a^3_B}$, $\Psi_{ns}(0) =
1/\sqrt{\pi n^3a^3_B}$, $E_{1s} = - \alpha/2 a_B$ and $E_{ns} = -
\alpha/2a_B n^2$ we get
\begin{eqnarray}\label{label6.6}
\delta \epsilon^{\rm (ds)}_{1s} = -
\frac{8}{9}\,\alpha^4\,\mu^3\,(2a^{1/2}_0 +
a^{3/2}_0)^2\,\sum^{\infty}_{n = 2}\frac{1}{n(n^2 - 1)} = -
\frac{2}{9}\,\alpha^4\,\mu^3\,(2a^{1/2}_0 + a^{3/2}_0)^2.
\end{eqnarray}
The contribution of the continuous spectrum of the $\pi^-p$ pair,
coupled by the attractive Coulomb field in the S--wave state, can be
determined by
\begin{eqnarray}\label{label6.7}
\delta \epsilon^{\rm (cs)}_{1s} =
\frac{4\pi^2}{9}\,\frac{1}{\mu^2}\,(2a^{1/2}_0 +
a^{3/2}_0)^2\,|\Psi_{1s}(0)|^2 \int^{\infty}_0dE\,
\frac{|\Psi_E(0)|^2}{E_{1s} - E},
\end{eqnarray}
where the abbreviation (cs) means the {\it continuous spectrum}.  The
wave function $\Psi_E(\vec{r}\,)$ of the $\pi^-p$ pair, coupled by the
attractive Coulomb fields in the S--wave state with a continuous
energy spectrum, is equal to \cite{LL65,LL65a}
\begin{eqnarray}\label{label6.8}
\Psi_E(\vec{r}\,) = \sqrt{\frac{1}{4\pi}\,\frac{\alpha
\mu^2}{\displaystyle 1 - e^{\textstyle\,-2\pi\alpha \mu/k}}}\,F\Big(1
+ i\,\frac{\alpha \mu}{k}, 2, 2ik r\Big),
\end{eqnarray}
where $k = \sqrt{2 \mu E}$ and $F(a,b, z)$ is a confluent
hypergeometric function \cite{MA72}. The wave function
$\Psi_E(\vec{r}\,)$ is normalized by
\begin{eqnarray}\label{label6.9}
\int d^3x\,\Psi^*_{E\,'\,}(\vec{r}\,) \Psi_E(\vec{r}\,) = \delta(E\,'
- E).
\end{eqnarray}
At $\vec{r} = 0$ we get \cite{LL65,LL65a}
\begin{eqnarray}\label{label6.10}
|\Psi_E(0)|^2 = \frac{1}{4 \pi }\,\frac{\alpha \mu^2}{\displaystyle 1
- e^{\textstyle\,-\,2\pi\alpha \mu/k}},
\end{eqnarray}
Substituting (\ref{label6.10}) into (\ref{label6.7}) and changing
variables $E \to k^2/2\mu$ we obtain
\begin{eqnarray}\label{label6.11}
\delta \epsilon^{\rm (cs)}_{1s} =
-\,\frac{2\pi}{9}\,\alpha\,(2a^{1/2}_0 +
a^{3/2}_0)^2\,|\Psi_{1s}(0)|^2 \int^{\infty}_0\frac{dk\,k}{k^2 +
\alpha^2 \mu^2}\,\frac{1}{\displaystyle 1 -
e^{\textstyle\,-\,2\pi\alpha \mu/k}}.
\end{eqnarray}
The integral over $k$ is divergent. As has been shown in \cite{AI03}
it should be regularized by a cut--off $K = \alpha \mu$. A divergent
part can be removed by a renormalization of the reduced mass of the
$\pi^-p$ pair. The regularized contribution of the continuous spectrum
to the shift of the energy level of the ground state of pionic
hydrogen reads
\begin{eqnarray}\label{label6.12}
\delta \epsilon^{\rm (cs)}_{1s} =
-\,\frac{2\pi}{9}\,\alpha\,(2a^{1/2}_0 +
a^{3/2}_0)^2\,|\Psi_{1s}(0)|^2 \int^{\alpha \mu}_0\frac{dk\,k}{k^2 +
\alpha^2 \mu^2}\,\frac{1}{\displaystyle 1 -
e^{\textstyle\,-\,2\pi\alpha \mu/k}}.
\end{eqnarray}
Dropping the contribution of the exponential, which is insignificant
in the physical region of relative momenta\,\footnote{A maximum value
the exponential acquires at the upper limit. Setting $k = \alpha \mu$
one gets $e^{\textstyle\,-\,2\pi\alpha \mu/k} = 1.87\times 10^{-3}$
that is less than 0.2$\%$.}, we get
\begin{eqnarray}\label{label6.13}
\delta \epsilon^{\rm (cs)}_{1s} = - \,\frac{\pi}{9}\,\alpha\,{\ell
n}2\,(2a^{1/2}_0 + a^{3/2}_0)^2\,|\Psi_{1s}(0)|^2 = - \frac{{\ell
n}2}{9}\,\alpha^4\,\mu^3\,(2a^{1/2}_0 + a^{3/2}_0)^2.
\end{eqnarray}
Thus, the total correction $\delta \epsilon_{1s} = \delta
\epsilon^{\rm (ds)}_{1s} + \delta \epsilon^{\rm (cs)}_{1s}$ to the
shift of the energy level of the ground state, caused by the $ns \to
1s$ transitions and the continuous spectrum of the $\pi^-p$ pair
coupled by the attractive Coulomb field in the S--wave state, is equal
to
\begin{eqnarray}\label{label6.14}
\delta \epsilon_{1s} &=& - \,\frac{2}{9}\,\Big(1 +
\frac{1}{2}\,{\ell n}2\Big)\,\alpha\,(2a^{1/2}_0 +
a^{3/2}_0)^2\,|\Psi_{1s}(0)|^2 =\nonumber\\ &=& - \frac{2}{9}\,\Big(1
+ \frac{1}{2}\,{\ell n}2\Big)\,\alpha^4\,\mu^3\,(2a^{1/2}_0 +
a^{3/2}_0)^2.
\end{eqnarray}
Comparing $\delta \epsilon_{1s}$ with the DGBT formula we obtain
\begin{eqnarray}\label{label6.15}
\frac{\delta \epsilon_{1s}}{\epsilon_{1s}} = \frac{\alpha}{3}\,\Big(1
+ \frac{1}{2}\,{\ell n}2\Big)\,\mu\,(2a^{1/2}_0 + a^{3/2}_0) =
0.76\times 10^{-3} = 0.076\,\%,
\end{eqnarray}
where we have used the experimental values of the S--wave scattering
lengths $a^{1/2}_0$ and $a^{3/2}_0$ \cite{PSI2}
\begin{eqnarray}\label{label6.16}
a^{1/2}_0 &=& (+ 0.1788 \pm 0.0043)\,m^{-1}_{\pi^-},\nonumber\\
a^{3/2}_0 &=& (- 0.0927 \pm 0.0085)\,m^{-1}_{\pi^-},\nonumber\\ 2
a^{1/2}_0 + a^{3/2}_0 &=& (+ 0.2649 \pm 0.0121)\,m^{-1}_{\pi^-}.
\end{eqnarray}
As has been calculated in \cite{AI03} the contribution to the shift of
the energy level of the ground state to the second order in strong
low--energy interactions $\epsilon^{(2)}_{1s}$ relative to the DGBT
result is equal to
\begin{eqnarray}\label{label6.17}
\delta^{(2)}_{1s} = \frac{\epsilon^{(2)}_{1s}}{\epsilon_{1s}} =
\frac{2\alpha}{\pi}\,\mu\,\frac{2(a^{1/2}_0)^2 + (a^{3/2}_0)^2}{2
a^{1/2}_0 + a^{3/2}_0} = (1.11\pm 0.06)\times 10^{-3} = (0.111 \pm
0.006)\,\%.
\end{eqnarray}
Therefore, a total contribution to the shift of the energy level of
the ground state of pionic hydrogen caused by strong low--energy
interactions makes up $(0.187\pm 0.007)\,\%$. Hence, it does not exceed
the experimental accuracy $0.2\,\%$ of the new set of experiments by
the PSI Collaboration \cite{PSI1}.

\section{Conclusion}

Within a quantum field theoretic, relativistic covariant and
model--independent approach we have derived the energy level
displacement of the excited $n\ell$ state of pionic hydrogen in terms
of the partial--wave scattering lengths of $\pi N$ scattering. We have
given the explicit calculation of the energy level displacements of
the excited $np$ states in terms of the P--wave scattering lengths of
$\pi^-p$ scattering. The shift of the energy level of the excited $np$
state is found in analytical agreement with the result obtained by
Ericson and Weise \cite{TE88}.  We have shown that the contributions
of the energy level displacements of the $np$ states to the
transitions $np \to 1s$ are much less than the experimental accuracy
of $0.2\%$ \cite{PSI1}. Therefore, they can be neglected for the
extraction of the experimental value of the energy level displacement
of the ground state of pionic hydrogen from the $np \to 1s$
transitions.

We have given numerical values only for the energy level displacements
of the excited $np$ states. These numerical values are needed for the
theoretical elaboration of experimental data on the $np \to 1s$
transitions of pionic hydrogen, used by the PSI Collaboration for the
measurements of the energy level displacement of the ground state of
pionic hydrogen \cite{PSI1}. Experimentally there are measured only
scattering lengths of $\pi N$ scattering in the P$\,(\ell = 1)$,
D$\,(\ell = 2)$ and F$\,(\ell = 3)$ wave states \cite{MN79}. The D--
and F--wave scattering lengths are by factors of 10 and 100,
respectively, smaller compared with the P--wave scattering lengths
\cite{MN79}. Hence, the energy level displacements of the excited
$n\ell$ states for $\ell \ge 2$ are negligible smaller compared with
the energy level displacements of the excited $np$ states. It is
obvious that for the contemporary level of accuracy of experimental
technique the energy level displacements of the excited $n\ell$ states
with $\ell \ge 1$ cannot be practically measured, and the
contributions of them should be neglected for the extraction of the
experimental value of the energy level displacement of the ground
state of pionic hydrogen from the $n\ell \to 1s$ transitions.

We have calculated the contribution of the $ns \to 1s$ transitions and
the continuous spectrum of the $\pi^-p$ pair, coupled by the
attractive Coulomb field in the S--wave state, to the shift of the
energy level of the ground state of pionic hydrogen induced by strong
low--energy interactions. The numerical value of this contribution
relative to the DGBT expression makes up $0.076\,\%$. Taking into
account the result obtained in \cite{AI03} the total shift of the
energy level of the ground state of pionic hydrogen to the second
order of perturbation theory in strong low--energy interactions makes
up $0.187\,\%$. This does not exceed the experimental accuracy 0.2$\%$
of the new set of experiments on the energy level displacement of the
ground state of pionic hydrogen by the PSI Collaboration \cite{PSI1}.

The obtained results confirm that the contributions of QCD
isospin--breaking and electromagnetic interactions, calculated by
Gasser {\it et al.} \cite{JG02} (see also \cite{TE03}), are the most
important for the precise extraction of the S--wave scattering lengths
of $\pi N$ scattering from the experimental data on the energy level
displacement of the ground state of pionic hydrogen.

\section*{Acknowledgement}

We are grateful to our referee for useful comments and Torleif Ericson
for numerous helpful discussions and calling our attention to the
results obtained in references \cite{TE67,LA69} and \cite{TE03,TE88}.

\newpage

\section*{Appendix. Calculation of momentum integral in 
(\ref{label5.1})}

In this Appendix we give an explicit calculation of the momentum
integral in (\ref{label5.1}). The wave function $\Phi_{np}(k)$ is
defined by (\ref{label2.10}). Substituting (\ref{label2.10}) in the
momentum integral in (\ref{label5.1}) we get
$$
\int \frac{d^3k}{(2\pi)^3}\,\sqrt{\frac{m_{\pi^-}m_p}{
E_{\pi^-}(k)E_p(k)}}\,k\,\Phi_{np}(k) = 
$$
$$
= \frac{1}{\pi^{3/2}}
\int^{\infty}_0dr\,r^2\,R_{n1}(r)\,\int^{\infty}_0dk\,
\sqrt{\frac{m_{\pi^-}m_p}{
E_{\pi^-}(k)E_p(k)}}\,k^3\,j_1(kr).\eqno({\rm A}.1)
$$
For the spherical Bessel function $j_1(kr)$ we use the expression
\cite{MA72a}
$$
j_1(kr) = - \frac{d}{dr}\frac{1}{r}\Big(\frac{\sin
kr}{k^2}\Big).\eqno({\rm A}.2)
$$
The integration over $k$ we carry out in the limit $m_p \to
\infty$. This yields
$$
\int^{\infty}_0dk\, \sqrt{\frac{m_{\pi^-}m_p}{
E_{\pi^-}(k)E_p(k)}}\,k^3\,j_1(kr) = \int^{\infty}_0dk\,
\sqrt{\frac{m_{\pi^-}}{ E_{\pi^-}(k)}}\,k^3\,j_1(kr) =
$$
$$
= \sqrt{m_{\pi^-}}\frac{d}{dr}\frac{1}{r}\frac{d}{dr}
\int^{\infty}_0\frac{dk\,\cos(kr)}{\displaystyle (m^2_{\pi^-} +
k^2)^{1/4}} = m_{\pi^-}\,\frac{2^{1/4}\sqrt{\pi}}{
\Gamma(1/4)}\,\frac{d}{dr}\frac{1}{r}\frac{d}{dr}
(m_{\pi^-}r)^{-1/4}K_{1/4}(m_{\pi^-}r),\eqno({\rm A}.3)
$$
where we have used the formula \cite{MA72b}
$$
K_{\nu}(xz) = \Gamma\Big(\nu + \frac{1}{2}\Big)\,
\frac{~(2z)^{\nu}}{\sqrt{\pi}\,x^{\nu}} \int^{\infty}_0\frac{\cos(xt)
dt}{(t^2 + z^2)^{\nu + 1/2}}.\eqno({\rm A}.4)
$$
Due to the appearance of the McDonald function $K_{1/4}(m_{\pi^-}r)$
the integrand of the integral over $r$ is localized around $r \sim
1/m_{\pi^-}$. This allows to take the wave function $R_{n1}(r)$ equal
to
$$
R_{n1}(r) = r\;\frac{2}{3}\,\frac{\sqrt{n^2 -
1}}{n^{5/2}a^{5/2}_B}.\eqno({\rm A}.5)
$$
Substituting ({\rm A}.3) and ({\rm A}.5) in ({\rm A}.1) we get
$$
\int \frac{d^3k}{(2\pi)^3}\,\sqrt{\frac{m_{\pi^-}m_p}{
E_{\pi^-}(k)E_p(k)}}\,k\,\Phi_{np}(k) =
$$
$$
= \frac{1}{3\pi}\,\frac{\sqrt{n^2 -
1}}{n^{5/2}a^{5/2}_B}\frac{2^{5/4}}{
\Gamma(1/4)}\int^{\infty}_0dx\,x^3\,\frac{d}{dx}
\Big(\frac{1}{x}\Big(\frac{d}{dx} x^{-1/4}K_{1/4}(x)\Big),\eqno({\rm
A}.6)
$$
where $x = m_{\pi^-}r$. Using the relation \cite{MA72c}
$$
\frac{1}{x}\Big(\frac{d}{dx} x^{-1/4}K_{1/4}(x)\Big) = -
x^{-5/4}K_{5/4}(x)\eqno({\rm A}.7)
$$
we transform the integral over $x$ to the form
$$
\int^{\infty}_0dx\,x^3\,\frac{d}{dx} \Big(\frac{1}{x}\Big(\frac{d}{dx}
x^{-1/4}K_{1/4}(x)\Big) =
-\int^{\infty}_0dx\,x^3\,\frac{d}{dx}\Big(x^{-5/4}K_{5/4}(x)\Big) =
$$
$$
= 3\int^{\infty}_0dx\,x^{3/4}\,K_{5/4}(x) = 3\cdot
2^{-1/4}\,\Gamma\Big(\frac{3}{2}\Big)\,\Gamma\Big(\frac{1}{4}\Big),
\eqno({\rm A}.8)
$$
where we have used the formula \cite{MA72d}
$$
\int^{\infty}_0dx\,x^{\mu}\,K_{\nu}(x) = 2^{\mu
-1}\,\Gamma\Big(\frac{\mu + \nu + 1}{2}\Big)\,\Gamma\Big(\frac{\mu -
\nu + 1}{2}\Big).\eqno({\rm A}.9)
$$
Substituting ({\rm A}.8) in ({\rm A}.6) we get
$$
\int \frac{d^3k}{(2\pi)^3}\,\sqrt{\frac{m_{\pi^-}m_p}{
E_{\pi^-}(k)E_p(k)}}\,k\,\Phi_{np}(k) = \sqrt{\frac{n^2 - 1}{\pi n^5
a^5_B}}.\eqno({\rm A}.10)
$$
Using this result we calculate the energy level displacement of the
excited $np$ state of pionic hydrogen (see Section 5).


\newpage


\begin{thebibliography}{9}
\bibitem{PSI1} 
D. Gotto {\it et al.},
Physica Scripta, T {\bf 104}, 94 (2003).
\bibitem{PSI2}
H.--Ch. Schr\"oder {\it et al.},
Eur. Phys. J. C {\bf 21}, 473 (2001).
\bibitem{SD54} 
S. Deser, M. L. Goldberger, K. Baumann, and
W. Thirring, Phys. Rev. {\bf 96}, 774 (1954). 
\bibitem{TT61}
T. L. Trueman,
Nucl. Phys. {\bf 26}, 57 (1961).
\bibitem{TE67}
A. Partensky and M. Ericson,
Nucl. Phys. B {\bf 1}, 382 (1967).
\bibitem{LA69}
E. Lambert,
Helv. Phys. Acta {\bf 42}, 667 (1969).
\bibitem{DE76}
A. Deloff, Phys. Rev. {\bf C 13}, 730
(1976).
\bibitem{TE03}
B. Loiseau, T. E. O. Ericson, and A. W. Tomas,
PiN Newslett. {\bf 15}, 162 (1999), hep--ph/0002056;
Nucl. Phys. A {\bf 663}, 541 (2000), hep-ph/9907433; 
Nucl. Phys. A {\bf 684}, 380 (2001);
T. E. O. Ericson, B. Loiseau, and A. W. Thomas,
Phys. Rev. C {\bf 66}, 014005 (2002),  hep--ph/0009312;
T. E. O. Ericson, B. Loiseau, and S. Wycech,
Nucl. Phys. A {\bf 721}, 653c (2003), hep-ph/0211433;
T. E. O. Ericson, B. Loiseau, and S. Wycech,
{\it Determination of the $\pi^-p$ scattering length from pionic 
hydrogen}, Invited talk at Workshop on 
{\it Hadatom03} at ECT$^*$ in Trento, 12--18 October 2003, Italy; 
hep--ph/030134.
\bibitem{TE88}
T. E. O. Ericson and W. Weise,
in {\it PIONS AND NUCLEI}, Clarendon Press, 
Oxford, 1988, pp.192--198.
\bibitem{KH02}
K. Hagiwara {\it et al.},
Phys. Rev. {\bf D 66}, 010001 (2002).
\bibitem{AI03} 
A. N. Ivanov, M. Faber, A. Hirtl, J. Marton, and
N. I. Troitskaya, {\it On pionic hydrogen. Quantum field theoretic,
relativistic covariant and model--independent approach},
nucl--th/0306047 (to appear in EPJA).
\bibitem{JG02}
J. Gasser, M. A. Ivanov, E. Lipartia, M.
Moj$\check{\rm z}$i$\check{\rm s}$, and A. Rusetsky, Eur. Phys. J. C
{\bf 26}, 13 (2002).
\bibitem{JG99}
J. Gasser,
{\it CHIRAL PERTURBATION THEORY}, 
Nucl. Phys. Proc. Suppl. {\bf 86}, 257 (2000); 
Invited talk given at High Energy Physics International 
Euroconference on Quantum Chromo Dynamics - QCD '99, 
Montpellier, France, 7-13 Jul 1999; hep-ph/9912548 
and references therein.
\bibitem{CD03} 
A. N. Ivanov, M. Faber, M. Cargnelli, A. Hirtl, J. Marton,
N. I. Troitskaya, and J. Zmeskal,
{\it On pionic and kaonic hydrogen}, Proceedings of the Workshop 
on {\it CHIRAL DYNAMICS} at University of Bonn, 8--13 September 2003,
Germany,  p.127, hep--ph/0311212.
\bibitem{TE03a}
T. E. O. Ericson (private communication).
\bibitem{HP73}
V. De Alfaro, S. Fubini, G. Furlan, and C. Rossetti,
in {\it CURRENTS IN HADRON PHYSICS},
North--Holland Publishing Co., Amsterdam $\,\bullet\,$ London,
American Elsevier Publishing Co., Inc.,
New York, 1973.
\bibitem{LL65}
L. D. Landau and E. M. Lifshitz,
in {\it QUANTUM MECHANICS}, Volume 3 of Course of Theoretical
Physics, Pergamon Press, Oxford, 1965, pp.116--128.
\bibitem{SS61}
S. S. Schweber,
in {\it AN INTRODUCTION TO RELATIVISTIC QUANTUM FIELD THEORY},
Row, Peterson and Co$\,\bullet\,$ Evanston, Ill.,
Elmsford, New York, 1961.
\bibitem{MA72}
{\it HANDBOOK OF MATHEMATICAL FUNCTIONS WITH Formulas, 
Graphs, and Mathematical Tables}, edited by M. Abramowitz 
and I. A. Stegun, National Bureau of Standards, Applied 
Mathematics Series $\,\bullet\,$ 55, 1972.
\bibitem{SG67}
S. Gaziorowicz,
in {\it ELEMENTARY PARTICLE PHYSICS},
John $\&$ Sons, Inc., New York, 1967.
\bibitem{MN79}
M. M. Nagels {\it et al.},
Nucl. Phys. B {\bf 147}, 189 (1979).
\bibitem{JS03} 
J. Schweizer, 
{\it Hadronic atoms}, 
{\it On pionic and kaonic hydrogen}, Proceedings of the Workshop 
on {\it CHIRAL DYNAMICS} at University of Bonn, 8--13 September, 
Germany, 2003, p.38, hep--ph/0311212.
\bibitem{LL65a}
L. D. Landau and E. M. Lifshitz,
in {\it QUANTUM MECHANICS}, Volume 3 of Course of Theoretical
Physics, Pergamon Press, Oxford, 1965, pp.519--523.
\bibitem{MA72a}
(see \cite{MA72} p.439 formula (10.1.25)).
\bibitem{MA72b}
(see \cite{MA72} p.376 formula (9.6.25)).
\bibitem{MA72c}
(see \cite{MA72} p.376 formula (9.6.28)).
\bibitem{MA72d}
(see \cite{MA72} p.486 formula (11.4.22)).
\end{thebibliography}
\end{document}